\documentclass[12pt]{article}

\usepackage{amssymb,graphicx}

\usepackage{epsf}
\usepackage{graphicx,epsfig}
\usepackage{amsfonts}
\usepackage{amssymb}

\makeatletter
\renewcommand\section{\@startsection {section}{1}{\z@}%
                                 {-3.5ex \@plus -1ex \@minus -.2ex}
                                   {2.3ex \@plus.2ex}%
                                   {\normalfont\large\bfseries}}
\renewcommand\subsection{\@startsection{subsection}{2}{\z@}%
                                   {-3.25ex\@plus -1ex \@minus -.2ex}%
                                     {1.5ex \@plus .2ex}%
                                     {\normalfont\bfseries}}
\renewcommand\subsubsection{\@startsection{subsubsection}{3}{\z@}%
                                   {-3.25ex\@plus -1ex \@minus -.2ex}%
                                     {1.5ex \@plus .2ex}%
                                     {\normalfont\itshape}}
\makeatother

\newcommand{\Letter}{
\setlength{\textwidth}{16.5cm}
   \setlength{\textheight}{22.6cm}
    \hoffset=-0.5in
\voffset=-2.1cm }

\Letter

\begin{document}
\newcommand{\be}{\begin{equation}}
\newcommand{\ee}{\end{equation}}
\newcommand{\bea}{\begin{eqnarray}}
\newcommand{\eea}{\end{eqnarray}}
\newcommand{\barr}{\begin{array}}
\newcommand{\earr}{\end{array}}
\newcommand{\myfigure}[2]{\centerline{\resizebox{#1}{!}{\includegraphics{#2}}}}

\thispagestyle{empty}

\vspace*{0.3in}

\begin{center}
{\Large \bf The Eternal Sunshine of the Spotless Mind}

\vspace*{0.5in} {Richard Easther and Eugene A. Lim}
\\[.3in]
{\em Department of Physics, Yale University, New Haven CT 06520, USA } \\[0.3in]
\end{center}

\begin{center}
{\bf
Abstract}
\end{center}
\noindent
We extend the worldline measure for pocket formation in eternal inflation to allow for time-ordered bubble formation. Such a time-ordering is equivalent to imposing a preferred time-slicing on the ``parent'' de~Sitter space.   Using this measure, we describe a covariant version of the youngness paradox and show that the youngness paradox is a  gauge artifact if the parent spacetime is an unbroken de Sitter space, due to the lack of an explicit time-ordering for the bubble nucleation events.  We then show that one can add a ``clock'' to the de Sitter space, in the form of a vector field with a spontaneously broken symmetry that defines a unique timelike direction accessible to all observers.   Once this is done, the existence of a preferred slicing means that  the youngness paradox cannot be easily resolved. We use this to elucidate the apparent ``persistence of memory'' discussed recently by Garriga, Guth and Vilenkin, for inflationary universes produced by bubble nucleation.

\vfill
\newpage
 
\setcounter{page}{1}


\section{Introduction} \label{sect:introduction}

A key test for any inflationary model is whether it  makes a graceful transition from the accelerated inflationary phase to the dense, thermalized regime that sets up the initial conditions for a hot big bang.   This transition need not occur globally, but can be implemented successfully by a {\em local\/} transition to decelerated expansion, 
 In this case   each local transition to non-inflationary growth yields a {\em pocket universe\/}, which are perpetually separated from one other by an inflationary region.  These pockets can form after an individual Hubble volume evolves into a region of the potential where the semi-classical evolution dominates the stochastic evolution of the field \cite{Linde:1986fd,Goncharov:1987ir}, or via Coleman-De Luccia tunneling from a classically stable local minimum of the potential -- which was actually envisaged in Guth's original proposal  \cite{Guth:1980zm}.     In both situations, the comoving inflationary fraction of any given patch  will decrease continuously.  However, since the inflating region is growing exponentially, the physical inflationary volume increases without limit so from a global perspective inflation is future-eternal.  Moreover, once a pocket has formed one typically requires a further period of regular, slow roll inflation if its interior is to resemble the observable universe.

Obviously, a necessary condition for eternal inflation is the existence of either meta-stable minima in the potential, or regions where the potential is large enough and flat enough so that the stochastic evolution dominates the semi-classical rolling.   Given the vast complexity of the proposed string landscape, these conditions are likely to be satisfied at many points within this multi-dimensional potential, and eternal inflation is an almost inevitable corollary of the existence of the string landscape \cite{Bousso:2000xa,Feng:2000if}, providing significant new motivation for studies of this topic.  

Unfortunately,  the infinite spatial volumes implied by stochastic inflation -- along with the absence of a preferred time-slicing in inhomogeneous general relativistic cosmologies -- ensure that attempts to quantify the likelihood associated with the production of different types of pocket contain infinite numerical factors. These factors turn out to be strongly dependent on one's choice of slicing, and must be regularized if we are to make  unambiguous predictions about the mixture of pockets produced by the landscape, or other models of eternal inflation.  In \cite{Easther:2005wi} significant progress was made towards solving this dilemma by constructing a measure which did not explicitly depend on one's choice of slicing, and only involved quantities which were manifestly covariant. 
This so-called pocket counting measure was further extended to recycling models \cite{Aguirre:2006ak}, that is  models where an infinite sequence of bubbles can be nested inside of one another. This formalism was shown to be equivalent to the large spheres of Vilenkin and collaborators in \cite{Garriga:2005av}. In this approach, which we review in Section \ref{sect:measure}, the pockets are counted by a finite set of manifestly covariant worldlines. Similar world-line based approaches have also been proposed in   \cite{Bousso:2006ge,Bousso:2006ev}.

One  long-standing paradox of eternal inflation is the \emph{youngness paradox}, first suggested by Guth \cite{Guth:2000ka} based on the synchronous gauge probability measure devised initially by Linde, Linde and Mezhlumian \cite{Linde:1993xx}.\footnote{Note that the use of an explicit gauge choice by a candidate measure for eternal inflation does not guarantee that it will manifest some version of the youngness paradox. For specific examples see  \cite{GarciaBellido:1993wn,Linde:2007nm}.} The crux of this problem is that at any given point in time (for some choice of time coordinate $t$), the universe seems to be populated by an exponentially greater number of young pocket universes than old ones. This follows because the physical volume of de Sitter space is growing exponentially, while the rate at which physical volume is converted into pocket universes is typically time-independent.  Consequently, the number of pockets produced per unit time appears to increase exponentially, and young pockets vastly outnumber the old ones.  Conversely, since our universe is about 14 billion years old, it is apparently one of the very special old pockets \cite{Guth:2000ka,Linde:1993xx,Linde:1994gy}.

This paradox is very difficult to disentangle from  the gauge-dependence of time-slicings in the eternally inflating universe (see  \cite{Guth:2007ng} for a recent discussion).   Consequently, it seems logical to use the tools provided by worldline based measures of eternal inflation to investigate the youngness paradox. We thus begin our analysis by extending the worldline measure of \cite{Easther:2005wi} to include an explicit time-ordering (Section \ref{sect:measure}).   In doing so we appeal to time-invariance to show explicitly that, in a genuinely unbroken de~Sitter space, the youngness paradox appears when one chooses  {\em any\/} specific slicing.   However, while individual bubbles can and will collide with their neighbours,  two arbitrary bubbles will generically never interact with one another. In this case, there is always a spacelike interval between any pair of points chosen from the two bubble walls, and one cannot unambiguously time-order their creation. Consequently, one can always choose slicings in which any {\em given\/} pocket will appear to be old.   

This argument hinges on the absence of a universal clock, which can be read by a purely local measurement by an observer inside the parent de~Sitter space. In this case one has a natural time-slicing, and the youngness paradox returns with considerable force.  We explore this possibility by considering two ways in which ``hair'' can be attached to the de~Sitter space (Section \ref{sect:addinghair}).   The first is to explicitly add extra matter or gradient terms. Asymptotically, these will be subject to the usual de~Sitter space no-hair theorems \cite{Gibbons:1977mu}, and we argue that they will be diluted to a level at which they will be indistinguishable from quantum fluctuations after some {\em finite\/} time. Consequently, this type of perturbation does not provide a universal clock, and the youngness paradox is once again tamed.   Alternatively, one can introduce hair which does \emph{not} get erased as the universe expands; a situation we will refer to as \emph{grain}.  One way to do this is to violate Lorentz invariance by introducing a vector field with a  spontaneously broken symmetry. If this field's vacuum state has a vanishing energy, it does not modify the dynamics but provides a grain to the spacetime which picks out a preferred slicing. Consequently, the broken symmetry provides a clock which then propagates a unique slicing into the arbitrarily far future.  By construction, bubbles generated from this background have a physically meaningful formation time, and the youngness paradox must be taken seriously. 

With this construction in hand, we then elucidate a puzzling aspect of recent work by Garriga, Guth and Vilenkin \cite{Garriga:2006hw}, who showed that bubbles in a de~Sitter background apparently retain some memory of their creation. As a consequence, bubble collisions are anisotropic, defining a special center accessible to hypothetical observers inside each bubble. We show that the construction of \cite{Garriga:2006hw} implicitly defines a vector field which allows a clock to propagate in the background de~Sitter space, allowing us to understand why the youngness paradox rears its head in their construction. We then close by arguing that any model of eternal inflation where the parent space is time-translation invariant will be free of the youngness paradox, and that the apparent age of our own universe may thus be sufficient grounds to disfavor scenarios in which the youngness paradox is manifest.

\section{The Worldline Measure} \label{sect:measure}

\begin{figure}[t]
\myfigure{3in}{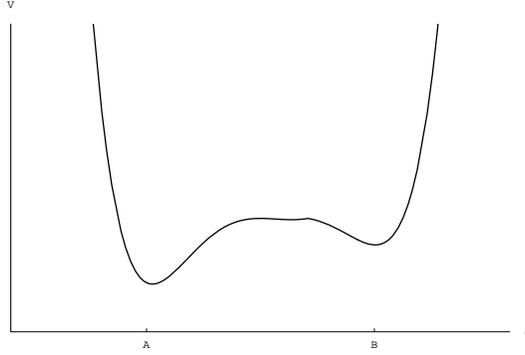}
\caption{A simple schematic potential for eternal inflation. The ``parent'' de~Sitter space has $\phi \sim 0$ and stochastic fluctuations produce pockets of type $A$ or $B$, where $V_A \ll V_B$. The inflaton begins at the little dimple between the two vacua and has a probability of tunneling into either the $A$ or $B$ which depends on the energy difference between the two, \emph{a la} Coleman-De~Luccia \cite{Coleman:1980aw}. }
\label{fig:2vacuaplot}
\end{figure}

We begin by describing the wordline measure \cite{Easther:2005wi}. For simplicity, we will use a specific model of eternal inflation, in which all the pockets are terminal vacua, so tunneling is a strictly one-way process.
 Consider a potential $V(\phi)$ with the form  shown in Figure \ref{fig:2vacuaplot}, where the inflaton begins on the hill in the center of the potential. At every timestep $\delta t$, on the order of the Hubble time, we divide an initial comoving patch into $4$ new Hubble patches.\footnote{Note we have projected out the third spatial dimension here, in order to visualize the formalism more clearly.} In each interval $\delta t$, an inflating Hubble patch may tunnel into vacuum $A$, tunnel into vacuum $B$, or continue inflating. We denote the first two probabilities $q_A$ and $q_B$, so the probability of not tunneling is simply $1-q_A-q_B$.  We assume that the universe thermalizes after tunneling to $A$ or $B$.

We start the process from $t=0$ and follow the system for $n$ timesteps, where $n$ is a very large number. We wish to compute the probability distribution of pocket universes $P_A$ and $P_B$. To do this, we lay down on the initial comoving surface a set of $N$ \emph{randomly distributed} worldlines. For simplicity,  we assume these worldlines are drawn from a flat distribution (with respect to spatial co-ordinates on the initial hypersurface), with zero velocities.

We follow each worldline along its trajectory until it intersects or ``tags'' a thermalized pocket. Once a pocket is tagged, any subsequent worldline which intersects it will be discarded, thus reducing the effective number of worldlines from $N$ to $N'$, where $N>N'$. We then count the total number of pockets thus tagged, and this number divided by $N'$ is the probability we seek
\begin{equation}
P_A = \frac{N_A}{N'}, \qquad P_B = \frac{N_B}{N'} = 1 - P_A \label{eqn:ELMmeasure}
\end{equation}
These expressions converge as we take $N\rightarrow \infty$ and, for the toy model here, $P_A\rightarrow q_A/(q_A+q_B)$ and $P_B\rightarrow q_B/(q_A+q_B)$.  

Since  worldlines which are manifestly covariant, the measure is gauge independent. Even so, we {\em have\/} introduced a gauge, in that our procedure implicitly assumes a slicing.  However, the time parameter $t$ is the worldlines' affine parameter, so it is  not an observable, and thus cannot enter into the final answer, Eqn. (\ref{eqn:ELMmeasure}). In other words, it does not really matter in what \emph{order} the worldlines each individually intersect pockets -- we simply need the final count of tagged pockets. 

We  now extend the worldline measure so we can examine the number of number of new pockets being formed at any given timestep.  In this case, the pocket is now tagged with a number corresponding to the first timestep in which one of our congruence of worldlines intersects its boundary.  This labeling is, of course,  slicing dependent.  We now write $N_A$ as 
\begin{equation}
N_A(n \delta t) = \sum_{i=1}^{n} \delta N_A(i \delta t)
\end{equation}
where $\delta N_A(i \delta t)$ is the number of pocket universes of type $A$ that are tagged for the first time during the $i$-th timestep. Similarly the probability of finding a pocket of type $A$ that was tagged during the $i$-th timestep is
\begin{equation}
P_A(i \delta t) = \frac{\delta N_A(i\delta t)}{N'}. \label{eqn:timemeasure}  
\end{equation}
We can recover our initial definition by summing up all the distributions to make it time-step independent
\begin{equation}
P_A =  \sum_{i=1}^{n} P_A(i \delta t) =  \frac{N_A}{N'} \, .
\end{equation}

As the parent de~Sitter space expands the number of members of our initial set of wordlines contained within a specific horizon volume will fall.  Consequently,   the number of pockets tagged in an interval $\delta t$ increases with time: hence any random drawing from this distribution of pockets is more likely to result in a young pocket than an old one. This is the youngness paradox cast in bubble counting language.

At early times, when a single pocket nucleates it absorbs a large number of individual worldlines, but we only count the pocket once.  Since the typical degeneracy decreases with $t$, $\delta N_A(i \delta t)$ will be a rising factor of $i$.  At very late times,  the initial set of worldlines will be diluted to the point where no degeneracies remain, and all of our worldlines exist inside of their own Hubble volumes.  In this limit, the number of surviving worldlines will follow a radioactive decay law, since a constant fraction of these worldlines will intersect pockets in a given interval if the nucleation rate per Hubble time is constant.     Provided $N$ is finite  these sums will be well-behaved as $n \rightarrow \infty$. At very late times  $\delta N_A(t) \rightarrow 0$, all worldlines will eventually intersect pockets and the rate at which new pockets are tagged will eventually approach zero. This formulation also converges in the limit $N \rightarrow \infty$ as shown in \cite{Easther:2005wi}.

\section{The Youngness Paradox and its Resolution} \label{sect:extendedELM}

This extended measure enjoys the same benefits as the original measure, as it is both covariant and convergent, even for multiple vacua. However, the specific choice of slicing associated with the time-step and initial hypersurface renders it gauge-dependent, and it appears that we now have a covariant version of the youngness paradox.  

\begin{figure}[t]
\myfigure{3in}{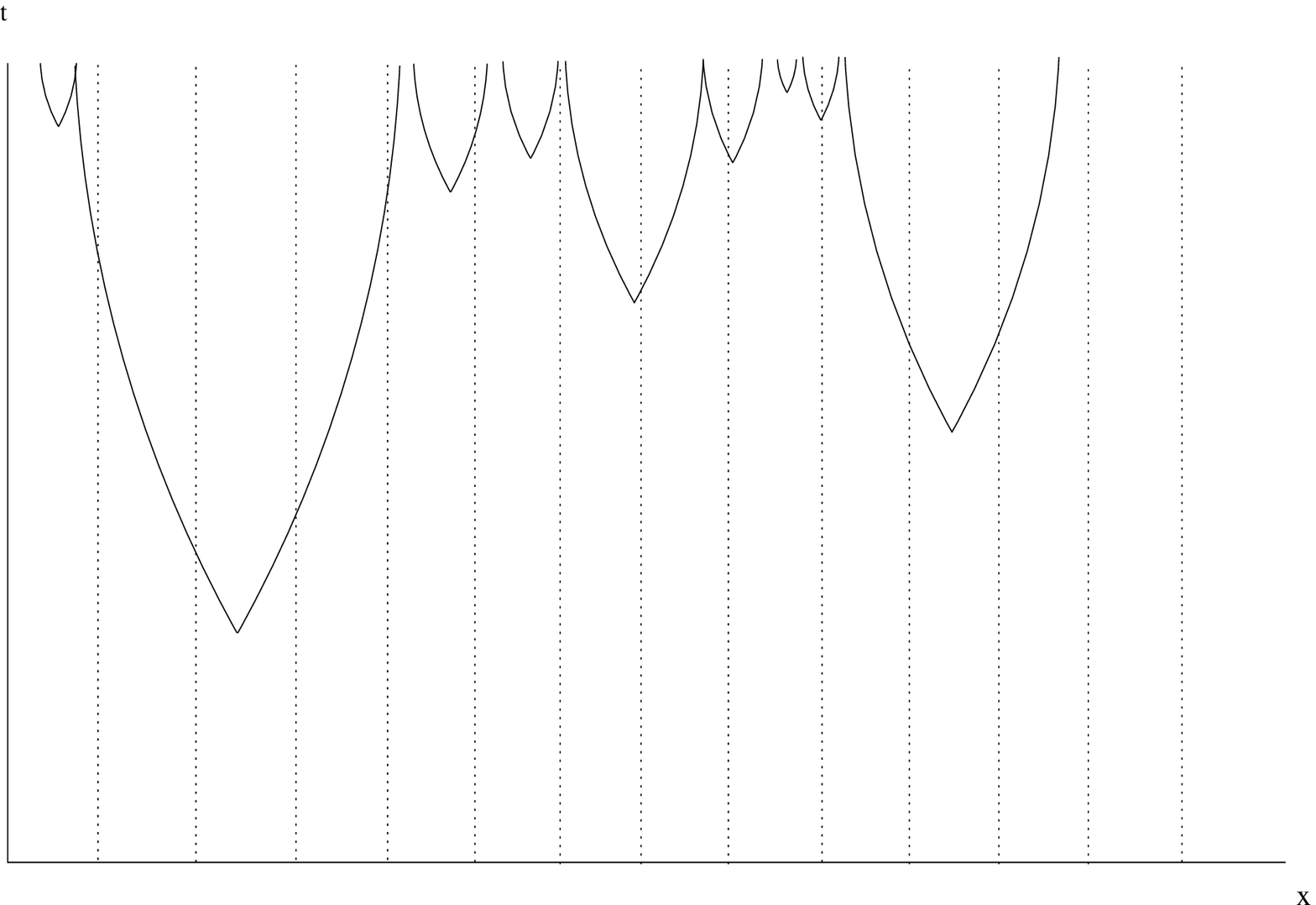}
\myfigure{3in}{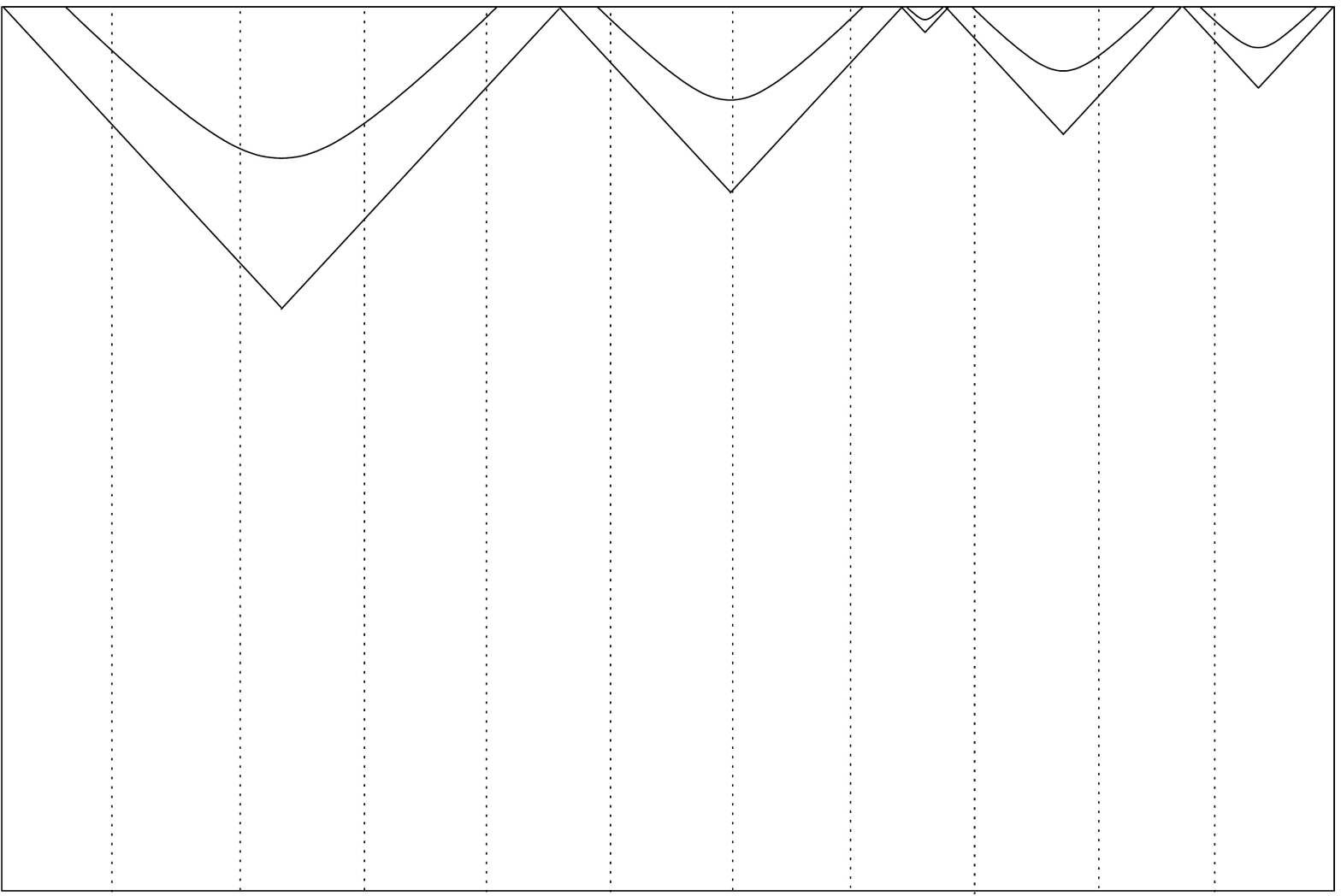}
\caption{Bubble formation in the comoving planar (top) and conformal (bottom) coordinates. The planar coordinates only cover half of the conformal space. Dotted lines denote static comoving observers in the respective coordinates; note that the trajectories of these two sets of static observers do not coincide. Both diagrams only show a small portion of the entire spacetime and we have assumed that no bubble collisions occur. }
\label{fig:comvsco}
\end{figure}

However, we can always choose a new time coordinate in such a way that as to ensure that a ``young'' pocket in one slicing  will be ``old'' in the new.  Note that this is not equivalent to boosting or changing the velocities of the worldlines; changing the observers' peculiar velocities relative to the background, does not change the probability distribution Eqn. (\ref{eqn:timemeasure}) since any initial velocities will quickly be redshifted away.

\begin{figure}[t]
\myfigure{5in}{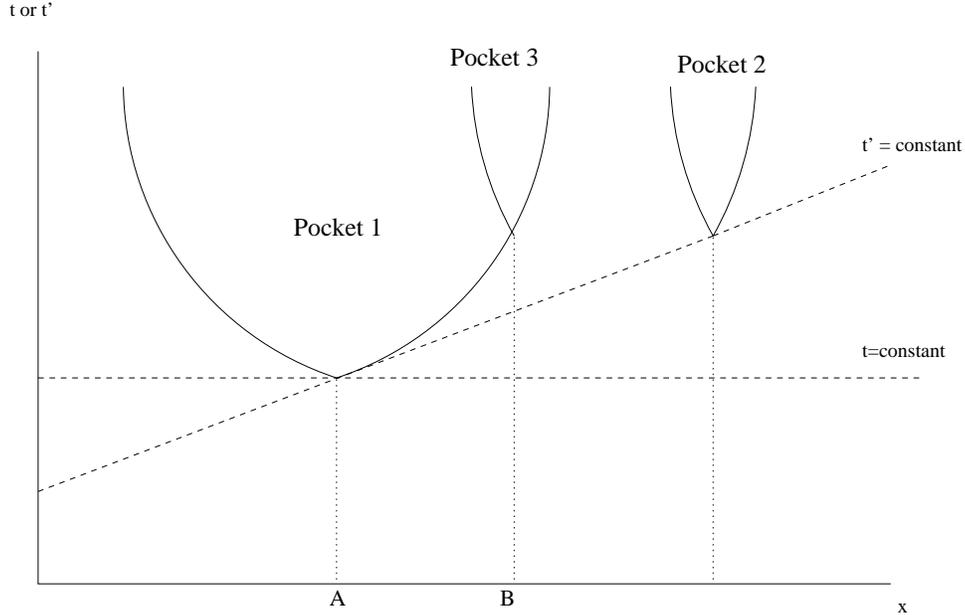}
\caption{A patch of 3 dimensional de~Sitter space in co-moving coordinates $(t,x,y)$ where we have suppressed the $y$ coordinate with each point on the diagram a 2-sphere. Also shown is a hyperslicing of constant $t'$ in the transformed coordinate system. In this coordinate system the point of nucleation is discontinuous (``pointy'') as one can show by computing $dt/dx \propto \pm e^{t_0}$. It is clear here that pocket 1 nucleate at a later $t$ than pocket 2, but in the $t'$ coordinate, they nucleate the same time. Also shown here are the tagging of pocket 1 by two different worldlines $A$ and $B$. Since the pocket wall is null, the points of entry for both $A$ and $B$ are exactly the same temporally and both observers (if they shall meet inside the pocket) will describe exactly the same evolution. In other words, for observer $B$, pockets $1$ and $3$ would be degenerate.}
\label{fig:comoving}
\end{figure}

To illustrate this, consider a small patch in 3 dimensional de~Sitter space $dS_3$. In co-moving coordinates, the metric is described by
\begin{equation}
ds^2 = -dt^2 + e^{2t}(dx^2+dy^2)
\end{equation}
where we have set the de~Sitter radius $H=1$ for simplicity. This coordinate system only covers the top-half of the entire de~Sitter space, but it suffices for our purposes (Figure \ref{fig:comvsco}). Once a pocket nucleates, the bubble wall travels at the speed of light \cite{Coleman:1980aw} and it thus traces out a null surface
\begin{equation}
ds^2 = 0 = -dt^2+e^{2t}(dx^2+dy^2). \label{eqn:dS3metric}
\end{equation}
Assuming that the nucleation occurs at point $(x_0,t_0)$,  the solution (where we have without loss of generality chosen the $y=0$ slice) is
\begin{equation}
x(t)-x_0 =\pm( e^{-t}- e^{-t_0})
\end{equation}
so, as $t\rightarrow \infty$, the bubble wall goes to a trajectory at fixed $x$, $x(t)-x_0 \rightarrow \pm e^{t_0}$. Note that bubbles which formed at later $t$ carve out a smaller co-moving slice in $x$, as shown in Figure \ref{fig:comvsco}. Consider now two pockets $1$ and $2$, nucleating at different points $(x_1,t_1)$ and $(x_2,t_2)$ where $t_2>t_1$, as in Figure (\ref{fig:comoving}). It is tempting to say that pocket $1$ nucleates earlier than $2$, and is thus less probable to be picked in any form of random selection,  given the probability measure Eqn. (\ref{eqn:timemeasure}).   However, performing the  simple coordinate transformation
\begin{equation}
t' = t+bx~,~x'=x \label{eqn:simpletransform}
\end{equation}
where $b$ is a dimensionless constant, the metric Eqn. (\ref{eqn:dS3metric}) becomes
\begin{equation}
ds^2 = -dt'^2 +2bdt'dx'+(e^{2t'}-b^2)dx'{}^2+e^{2t'}dy^2.
\end{equation}
This is still a de~Sitter space, given that geometry is independent of the coordinate choice, and one can directly verify that it solves the Einstein equations for a pure cosmological constant, $G_{\mu\nu}=-g_{\mu\nu}$. Moreover, both sets of co-ordinates cover the same patch of de~Sitter space, and constant $x$ trajectories still overlap. In the prime coordinates, we can find two bubbles which nucleate on the same slice, or
\begin{equation}
t'_1(x_1) = t'_2(x_2) \, .
\end{equation}
However, Eqn. (\ref{eqn:simpletransform}) gives
\begin{equation}
t_1 = t_2 + b(x_2-x_1).
\end{equation}
Thus it is clear that as long as $x_2\neq x_1$ we can choose  $b\neq 0$, such that we can reverse the apparent relative age of the bubbles.   

The analysis above recapitulates long established properties of bubble formation in  de~Sitter space. One can argue that the presence of nucleating bubbles will eventually break the symmetries of de~Sitter space by filing up the spacetime at future infinity.\footnote{We thank Alex Vilenkin for pointing this out ot us.}  However, we have labored this point to make it clear that the seemingly innocuous idea of classifying a pocket's \emph{age} by its time of nucleation is inherently ambiguous, if the parent space is an unbroken de~Sitter space. \emph{If} we have access to the full de~Sitter symmetries, then there is no agreement as to {\em which\/} of the pockets are young and which of the pockets are old, and the youngness paradox is reduced to the level of a curiosity with no explanatory or predictive power. Physically, we understand that this reflects the time-translation invariance of de~Sitter space.

\begin{figure}[t]
\myfigure{3in}{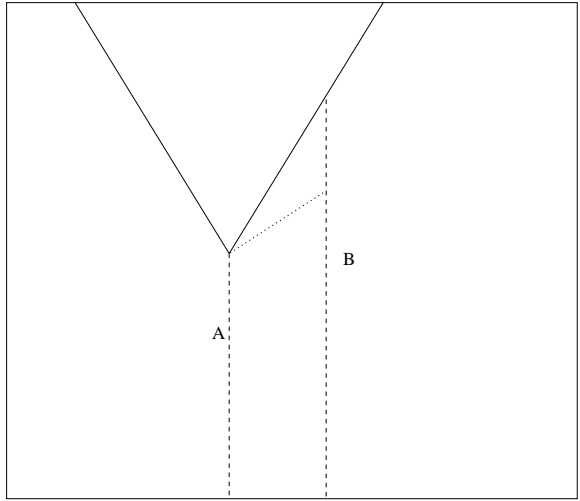}
\caption{The importance of the null-likeness of the bubble wall is illustrated here. If the bubble wall is timelike, then observer $A$ can send information to observer $B$ as he hits the bubble wall warning her of the impending wall ahead. If the signal he sends can survive a phase transition, he can even do that after he crosses into the bubble. This fact allow the two observers to time-order their arrival, and defines a center for the bubble. Also, if observer $B$ encounter a different bubble than $A$'s, then she can time-order the bubbles.}
\label{fig:nulltype}
\end{figure}

Furthermore, since the bubble wall is traced by null geodesics, no observer can see the wall before she is engulfed by it, and we cannot unambiguously define the point at which the bubble forms. Thus, any observer encountering a pocket can treat the point at which they enter it as the center, and there can be no difference between what an observer entering Pocket 1 at location $A$ or location $B$ would actually see.  So not only are the pockets not time-ordered relative to one another, we cannot even pick out a preferred point inside a single pocket by associating it with an actual origin where tunneling occurs.  Of course, if the bubble wall was timelike, then observer A could send a signal to observer B before Observer B falls into the same pocket, and the universe within the pocket would have a cleanly defined center (Figure \ref{fig:nulltype}).

We can extend this argument to the case of colliding bubbles. Imagine two observers existing happily inside separate bubbles which then collide at some spacetime point $P$. Although  our sketch suggests that Observer A's bubble formed before that of Observer B, we could reverse this illusion by reslicing the de~Sitter space.\footnote{If on the other hand an observer in a bubble experience a second decay into another lower-energy vacuum, i.e. the multiverse supports nested bubbles, then that hypothetical observer will see that the new spacetime has a lower total vacuum energy density than its predecessor.} Again, the null nature of the bubble wall plays an important role; if the wall is timelike, observer A can warn observer B ahead of her impending collision and send her information regarding the nature of the pocket she will be entering. If Observer B then falls into a different pocket (that then collides with Observer's A pocket) both observers will be able to compare notes inside the newly merged pocket and conclude they have different evolutionary histories.
 
\section{Adding  Hair and Adding Grain} \label{sect:addinghair}

In the previous section, we showed that the symmetries of de~Sitter space make it impossible to time-order pocket universes, and reminded the reader that, because  all points on the bubble wall are separated by light-like intervals, there is no preferred center.  However, it is not obvious that time translation invariance is a necessary condition for eternal inflation. We now seek mechanisms  to break the time-translation invariance of de~Sitter while preserving the existence of eternal inflation.  This is synonymous with adding hair to the de~Sitter space, and there are well-known theorems showing that a broad class of possible mechanisms for doing this only yield a local disturbance, and full de~Sitter invariance is restored asymptotically \cite{Hawking:1973uf}.  

For instance, we could imagine adding a clock to our de~Sitter space by postulating that it was created with a finite but homogeneous distribution of matter or radiation.  However,   the density of this matter or radiation will decrease exponentially, and quickly reach the point where there is less than one particle (or photon) per horizon volume, after which it can have no discernible effect on the dynamics, and full de~Sitter invariance is effectively restored. Moreover, note that for any finite amount of additional matter only a finite amount of time is needed to effectively remove all traces of this hair.  Alternatively, one could imagine adding a slowly-rolling scalar field whose potential makes a negligible contribution to the overall energy density, and use the field's value as a clock. Unfortunately this process will again fail, since the kinetic energy of the field will become vanishingly small at late times, and its expectation value will be dominated by quantum fluctuations (which are proportional to the $H$ of the de~Sitter background), rendering it useless as a clock.

On the other hand, we can imagine breaking the background Lorentz invariance while making no changes to the background dynamics. In this case, we could effectively give spacetime a wood-like ``grain'' which would define a preferred time-like direction.  Perhaps the simplest way is to set up a preferred frame picked out by some field; for example we may lay down a congruence of curves which define a vector field $V^{\mu}(x)$, and attach some observable meaning to it. In \cite{Garriga:2006hw}, as we will show below, such a congruence was defined to be orthogonal to some initial spacelike hypersurface and the vector field represents the direction of constant bubble nucleation rate. 

The addition of such eternally non-vanishing grain naturally provides  a clock with which we can time-order the formation of bubbles.  Now the ``formation time'' label Eqn. (\ref{eqn:timemeasure})  becomes physically meaningful, and the youngness paradox is a genuine problem.  For definiteness, let us consider a particularly simple model. In $3+1$ dimensions, using the planar metric and restoring the de~Sitter scale we have $H_{ds}$
\begin{equation}
ds^2 = -dt^2 + e^{2H_{ds}t}(dx^2+dy^2+dz^2). \label{eqn:dSmetric}
\end{equation}
Now consider a scalar field $\chi(x^{\mu})$ which permeates the universe. Furthermore, each observer in the universe is armed with a $\chi$-meter which allows them to measure the value of this field so that $\chi$ is a physical observable. For $\chi$ to function as a clock, it must increase monotonically in the direction of all possible time-like trajectories $b^{\mu}(\tau)$ \cite{Bobbook} where $\tau$ is the affine parameter,
\begin{equation}
g_{\mu\nu}\frac{db^{\mu}}{d\tau}V^{\nu}<0 \label{eqn:causalstability}
\end{equation}
and 
\begin{equation}
V^{\mu} \equiv \nabla^{\mu}\chi \label{eqn:vectordef}
\end{equation}
is the vector field   defined by the tangent vectors to the direction of the clock motion. As long as the norm is timelike, so that  $V^{\mu}V_{\mu}<0$, then indeed $\chi$ increases monotonically along all possible timelike trajectories. We choose a simple vector field with a constant timelike norm, so that its amplitude does not redshift away once we add dynamics
\begin{equation}
V^{\mu}V_{\mu} = -m^2 \label{eqn:constantnorm}
\end{equation}
where $m$ is some mass scale which defines the rate at which the clock ticks.  The vector field defines a congruence that picks out a preferred frame. To see this, consider the following ansatz for $V^{\mu}$
\begin{equation}
V^{\mu} = (m,0,0,0)  \label{eqn:vectorsolution}
\end{equation}
which using Eqn (\ref{eqn:vectordef}) we see that the trajectories of $\chi(x^{\mu})$ define a congruence of curves
\begin{equation}
x^{\mu} = (m(t+t_0),_0,y_0,z_0)
\end{equation}
labeled by initial conditions $(t_0,x_0,y_0,z_0)$, or 
\begin{equation}
\chi = m(t+t_0)+\vec{\mathbf{x}}_0.
\end{equation}
There is a shift symmetry $t\rightarrow t'+C$ which renders the clock ambiguous. In other words, the vector field is time-translation invariant, and hence observers can only measure $\chi$ up to an
unknown constant and thus cannot compare times. However, if we fix the value of $t_0$ for all curves, i.e. we construct the clock such that \emph{all} observers agree on the value of $t_0$ at all points in spacetime, then this ambiguity vanishes. We can do this by defining initial conditions for the value of $\chi$ at a Cauchy surface at $t=t_0$. Thus all observers can now agree on a common time convention. In the language of \cite{Garriga:2006hw}, this is the special initial hypersurface where there is no bubbles. One can argue that our construction is contrived, and we readily agree with this assessment. However it is not clear that this is a physical impossibility. If it is then $\chi$ becomes an unmeasurable quantity and provides us with an elegant way out of the youngness paradox\footnote{The presence of an ``initial'' hypersurface is certainly not a prerequisite for a physical universe, for example \cite{Aguirre:2003ck} argued that the universe could be past-eternal, and hence possess no cosmological singularity or beginning of time. The elliptic de~Sitter spacetime \cite{Parikh:2002py} that is their construction possess all the symmetry properties of regular de~Sitter space. We thank Anthony Aguirre and Jaume Garriga for a useful discussion and clarification of this point.}.
Before we describe a physical realization of such a clock, let us consider the ramifications of its presence. Clearly, we can now time-order the pockets, and each bubble has a definite nucleation time. We can define the nucleation time as the earliest time when the bubble first appears; in Figure \ref{fig:comoving}, this is simply the ``pointy'' spot. Moreover, for any given pocket, we can find the minimum value of $\chi$ for which it exists.  The corresponding spatial point is then the center of the bubble, and we can thus locate the nucleation point of a pocket. Inside the pocket, this center manifests itself as a preferred frame -- which accords with the understanding  that picking a global clock by laying down a vector field   breaks Lorentz invariance inside the pocket \cite{Jacobson:2000xp}. This opens up the intriguing possibility that an observer inside the bubble can see beyond the horizon wall and into the global physics of the multiverse, as first noted by the authors of reference \cite{Garriga:2006hw}.

What happens to this vector field after the bubble nucleates? There are several possibilities. One is simply that the phase transition leading to  bubble formation is so violent that the vector field itself becomes vanishingly small inside, and the bubble evolves as a plain vanilla FRW cosmology. If the vector field survives the phase transition,  inside the bubble it will still point in the same time direction as the parent de~Sitter space. Consequently, it breaks the translation symmetry, and will only preserve rotational symmetry about the central point. On the other hand, if this vector field contributes  a non-trivial energy density  it must satisfy the symmetries of the FRW bubble universe. In this case the vector field must then possess dynamics. 

Let us consider now an explicit example, namely  a vector field with a fixed time-like norm \cite{Kostelecky:1989jw,Jacobson:2000xp}
\begin{equation}
S = \int dx^4 \sqrt{g}\left(\frac{R}{16\pi G} + \Lambda- \frac{1}{4} \partial_{[\mu}V_{\nu]}\partial^{[\mu}V^{\nu]} + \alpha(V^{\mu}V_{\mu}+m^2) \right]
\end{equation}
which is guaranteed by  $\alpha$, a lagrange multiplier whose equation of motion is only solved if (\ref{eqn:constantnorm}) holds. Despite the presence of an additional dynamical field, the de~Sitter metric (\ref{eqn:dSmetric}) is still a solution to the Einstein equation \cite{Carroll:2004ai,Lim:2004js}
\begin{equation}
G_{\mu\nu} = - 8\pi G g_{\mu\nu}\Lambda.
\end{equation}
After nucleation the vector field reorients itself to the ``new'' time direction of the FRW bubble universe, but this vector field does not contribute to the expansion of the universe as its total energy density and pressure is always zero in the background.  Perturbations about this background are studied in \cite{Lim:2004js,Kanno:2006ty}, and its effects have been well-studied \cite{Eling:2004dk,Carroll:2005dj,Eling:2006ec,Eling:2006df,Foster:2005dk,Eling:2005zq,Heinicke:2005bp}.   We now see that $\chi$  provides a reliable clock for an eternally inflating universe.  Its initial conditions hypersurface breaks de~Sitter time-translational invariance globally, we can now use it to time-order the bubbles, and uniquely define their centers, giving the time used in   Eqn. (\ref{eqn:timemeasure}) a physical meaning.

Recently,   Garriga, Guth and Vilenkin  \cite{Garriga:2006hw} argued that bubble nucleation in de~Sitter space could implicitly define a preferred slicing.   In that paper, the authors postulated that there exists an initial spatial hypersurface at time $t_i$, in the coordinate system of the parent de~Sitter space, and that there are no bubbles everywhere along this slice. We can imagine that this is the beginning of eternal inflation\footnote{In \cite{Garriga:2006hw}, they define this surface as the time when past directed geodesics become incomplete \cite{Borde:2001nh}. While this provides the motivation to impose a surface of initial conditions at some finite time, one can imagine various other mechanisms to choose such a initial Cauchy surface to begin eternal inflation.}. Furthermore, they also assume that the bubble nucleation rate at any point in physical space is a constant. Now, consider an observer $O$ moving on a timelike trajectory inside the parent de~Sitter space which has not yet encountered a bubble wall. The probability $P$ of it encountering a bubble (or being hit by a bubble) as it moves between point $A$ and $B$ is then the determined by the spacetime volume between the past light cones of $A$ and $B$. Hence
\begin{equation}
\frac{d P}{d\tau} = \lambda \frac{d V_4}{d\tau}
\end{equation}
where $\tau$ is the affine parameter that parameterizes the trajectory, $\lambda$ is the nucleation rate per unit volume which is assumed to be constant, and $V_4$ is the $4$-volume of the past light cone
\begin{equation}
V_4 = \int\int dr dt e^{3t}r^2 dr dt. \label{eqn:4volume}
\end{equation}
Eqn. (\ref{eqn:4volume}) is bounded by the past light cone of the point in question, and cut off for values of $t$ earlier than that which defines the initial hypersurface. This is a finite quantity even when $t_i \rightarrow -\infty$. 

Consider the simplest example of a static observer with trajectory  
\begin{equation}
t(\tau) = \tau~,~r(\tau)=0, \label{eqn:statictraj}
\end{equation}
where $\tau$ is the affine parameter. It is an exercise in real analysis to integrate Eqn. (\ref{eqn:4volume}); we give the details in the Appendix. This leads to the solution
\begin{equation}
\frac{dP}{d t} =\lambda \frac{d V_4}{d t} =\lambda \frac{4\pi}{3}. \label{eqn:v4}
\end{equation}
Comparing this solution Eqn. (\ref{eqn:v4}) to the solution of the Lorentz violating vector field above\footnote{We  emphasise that while we have ensured the dynamical the breaking of Lorentz invariance by adding the field $V$ to the Lagrangian, this is not necessary. For example,  the CMB breaks Lorentz invariance in our universe by picking out a preferred constant temperature slicing.}
 Eqn. (\ref{eqn:vectorsolution}), we can identify $4\pi\lambda/3 \equiv m$. Hence the amount of 4-volume, from the nucleation point to the initial hypersurface measures the amount of time that has elapsed before the bubble nucleates, and allows us to time-order the bubbles. In other words, bubbles are more likely to nucleate at late times.

\section{Conclusion} \label{sect:conclusion}

We have extended the worldline measure to include the notion of time-ordering of bubble formation. Using this, we presented a covariant version of the youngness paradox, and resolved the paradox by showing that it is indeed a gauge issue.  However, the resolution of this paradox relies on two very important conditions. Firstly we make explicit use of the global time-translation invariance of de~Sitter space, and secondly the bubble wall has to be null. If either of these conditions are broken, then one has a physical clock that can time-order the bubble formations -- and in this case the youngness paradox is indeed a real problem and not simply a gauge issue.

To illustrate our points, we have considered a simple constant norm time-like vector field theory. This vector field defines a congruence in spacetime which serves as the preferred frame. The specifications of initial conditions on a Cauchy surface for this vector field globally breaks de~Sitter space, while preserving the local cosmological dynamics. We argue that this vector field acts like a clock, allowing local observers to check the cosmic time.

Once a clock has been specified,  each pocket now has a physical time label associated with it. Furthermore each pocket will possess an unambiguous center, breaking Lorentz invariance of the bubble universe. In particular,  if eternal inflation occurs in a background which admits an unambiguous global time-ordering of pocket formation, then we find that the youngness paradox must be taken seriously.   As a corollary, since our universe is observably very old, we might venture to infer that the parent de~Sitter space from which our pocket was born does was genuinely time-translation invariant.  In this light, it is interesting to return to the analysis of Garriga, Guth and Vilenkin \cite{Garriga:2006hw}, who conclude that the time invariance of de~Sitter space is broken by tunneling events, leading to an apparent ``persistence of memory''.   We believe that it is well-worth examining more complicated cases -- including pockets which are not themselves de~Sitter, nested bubbles, recycling vacua, and stochastic inflation rather than Coleman-de~Luccia tunneling -- in order to be sure this effect remains as the analysis is generalized.  Alternatively, we can seek out implementations of eternal inflation where the time invariance of the parent de~Sitter space is unbroken.

\section*{Acknowledgments}
We thanks Anthony Aguirre, Jaume Garriga, Alan Guth, Andrei Linde, Maulik Parikh and Alex Vilenkin for a sequence of extremely useful conversations about this work.  We also thank Matthew Martin for conversations during the early phase of this work.  
RE and EAL are supported in part by the United States Department of Energy, 
grant DE-FG02-92ER-40704.  This research was supported by grant RFP1-06-17 from The Foundational Questions Institute (fqxi.org)

\appendix
\section{Past light cone 4-volume for arbitrary trajectories} \label{appendix:integration}

In this appendix, we will slightly expand upon the derivation of $dV_4/d \tau$ done in \cite{Garriga:2006hw} for an observer traveling in arbitrary trajectories in de~Sitter.

\begin{figure}[t]
\myfigure{5in}{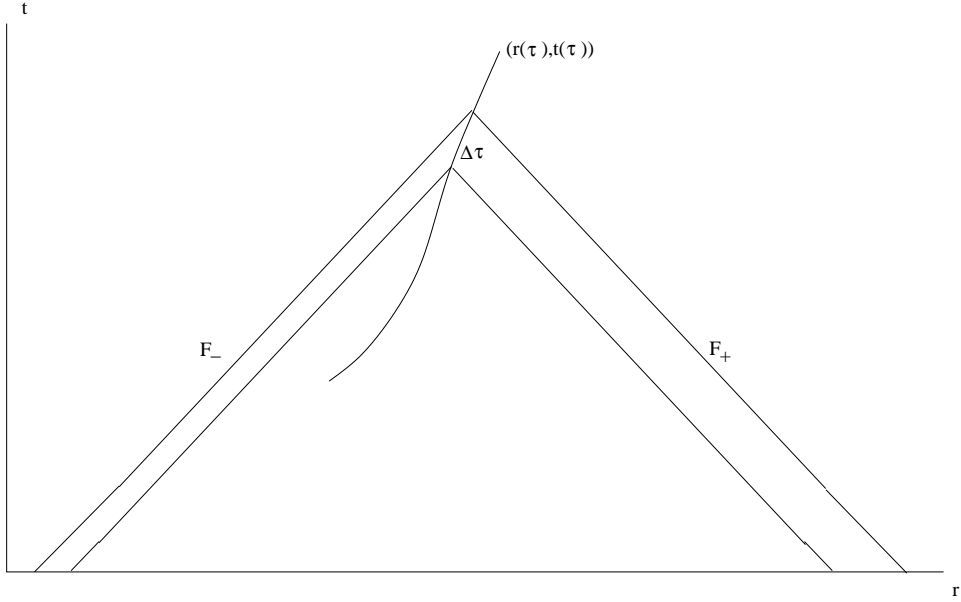}
\caption{The rate of change of the 4-volume enclosed by the past light cone of an observer traveling along the trajectory $(r(\tau),t(\tau))$ is shown in this figure. $F_+$ and $F_-$ are geodesics describing the look back boundary of the light cones. $\tau$ is the affine parameter of the curve; one can think of it as the proper time of the observer.}
\label{fig:4volume}
\end{figure}

We begin by considering an observer traveling along a trajectory parameterized by an affine parameter $\tau$ in de~Sitter space (Figure \ref{fig:4volume}). The four volume of the past light cone at each point in de~Sitter space using the planar coordinates and setting $H=1$ 
\begin{equation}
ds^2 = -dt^2 + e^{2t} (dr^2 + r^2d\Omega^2)
\end{equation}
is then
\begin{equation}
V_4 =  4\pi \int e^{3t}r^2 dt dr
\end{equation}
with the boundaries being the backward looking null cone to the initial hypersurface at $t_i$.

Using the fact that the backward light cone is symmetric under mirroring along the plane $r(\tau_0)$, we can split the integral into the $+$ and $-$ regions. As we will see, this  simplifies our computation
\begin{equation}
V_4 = 2\pi \int_+ e^{3t}r^2 dt dr + 2\pi \int_- e^{3t}r^2 dt dr \equiv  V_4^+ + V_4^-.
\end{equation}
Let us consider the $V_4^+$ integral first. It is convenient to make the following change of coordinates
\begin{equation}
a \equiv e^t~,~F_+ \equiv r-e^{-t} \label{eqn:plustransform}
\end{equation}
such that the integral now becomes (with Jacobian $J=1/a$)
\begin{equation}
V_4^+ = 2\pi\int \int^{a}_{a_i} (1+aF_+)^2 da dF_+.
\end{equation}
The reason we choose this coordinate system is that $F_+$ describes the trajectory of the $r>r_0$ light cone for any fixed value of $F_+$, and is thus independent of $a$. This means that we can do the $a$ integral first to obtain
\begin{equation}
V_4^+ = \left. 2\pi\int a+a^2F_+ + \frac{a^3 F_+^2}{3}  \right|^{a_0}_{a_i} dF_+.
\end{equation}
where $a_0$ is the intersection of the light cone described by $F_+$ with the trajectory of our observer. The $I_-$ integral can be similarly derived, using the following coordinate transformation
\begin{equation}
a \equiv e^t~,~F_- \equiv r+e^{-t}
\end{equation}
to be
\begin{equation}
V_4^- =  2\pi\int \int^{a_i}_{a_0} (1-aF_-)^2 dF_-=\left. 2 \pi \int a-a^2F_- + \frac{a^3 F_-^2}{3} \right|^{a_i}_{a_0} dF_-.
\end{equation}
Note that the limits on $a$ for the $I^-$ integral ranges from $a_0$ to $a_i$, i.e. we are integrating along the light cone boundary from $a_i$ back to $a_i$ via $a_0$. 

We have gone this far without explicitly stating the trajectory of the observer. To meaningfully proceed, let us choose the following trajectory
\begin{eqnarray}
t(\tau)& =& \tau \cosh \alpha \\ \label{eqn:traj_t}
r(\tau)&=& - e^{-\tau \cosh \alpha} \tanh \alpha. \label{eqn:traj_r}
\end{eqnarray}
where $\alpha$ is a constant. This corresponds to boosting the observer from $r=\mathrm{constant}$ trajectory to a velocity $dr/dt = a^{-1}\tanh \alpha$. This particular choice has the easy identification $dt = \cosh \alpha d\tau$, so we can immediately write down $dV_4/d\tau = \cosh \alpha (dV_4/dt) |_r=\mathrm{const} = 4\pi/3 \cosh \alpha$. But let us do the integral to show the operational steps to see how we can generalize this to arbitrary trajectories.

We now want to figure out the integration limits for $a$. The lower limit for $a$ is $a_i\rightarrow 0$ as we take $t_i\rightarrow -\infty$, while the upper limit is given by its intersection with the trajectory Eqs. (\ref{eqn:traj_t}) and (\ref{eqn:traj_r}) which in in terms of $F_+$ is then
\begin{equation}
a = \frac{-(1+\tanh \alpha)}{F_+}.
\end{equation}
The integral $I_+$ then becomes
\begin{equation}
V_4^+ =  \left . 2\pi\int  a-a^2F + \frac{a^3 F_-}{3} \right|^{-(1+\tanh \alpha)/F_+}_{0} dF_+= 2\pi \int -\frac{1+\tanh ^3 \alpha}{3F_+}  dF_+.
\end{equation}
But now using the transformation (\ref{eqn:plustransform}), and replacing $(t,r)$ with $\tau$
\begin{equation}
\frac{-dF_+}{F_+} = \cosh \alpha d\tau
\end{equation}
we get
\begin{equation}
\frac{dV_4^+}{d\tau} = \frac{2\pi}{3} \cosh \alpha (1+ \tanh^3 \alpha).
\end{equation}
We can go through similar steps for the $V_4^-$ term to get
\begin{equation}
\frac{dV_4^-}{d\tau} = \frac{2\pi}{3} \cosh \alpha (1- \tanh^3 \alpha).
\end{equation}
Notice that if we boost the observer to light speed going to the positive $r$ direction, $\alpha \rightarrow \infty$, the rate of 4-volume changes only for $V_4^+$ which makes sense.
The total rate of change of $V_4$ with respect to the affine parameter $\tau$ is then
\begin{equation}
\frac{dV_4}{d\tau} = \frac{4\pi}{3}\cosh \alpha
\end{equation}
as we expected, thus reproducing the computation done in \cite{Garriga:2006hw}. Setting $\alpha=0$, this means that the observer is static with respect to $r$, and thus we obtain Eqn. (\ref{eqn:v4}).

\end{document}